# Entropy-Engineered Synthesis of CuCo Nanometric Solid Solution Alloys for Near-100% Nitrate-to-Ammonia Selectivity

Yao Hu,[1] Haihui Lan,[2*] Bo Hu,[3] Jiaxuan Gong,[1] Donghui Wang,[1] Wen-Da Zhang,[1] Mo Yan,[4] Qi Wang,[1] Yulong Liu,[1] Huicong Xia,[5*] Mingde Yao,[6] Mingliang Du[1*]

[1] Key Laboratory of Synthetic and Biological Colloids, Ministry of Education, School of Chemical and Material Engineering, Jiangnan University, 214122, Jiangsu, China.

[2] Department of Chemistry, Massachusetts Institute of Technology, Cambridge, Massachusetts 02139, United States.

[3] College of Arts and Sciences, University of Illinois Urbana-Champaign, Champaign, IL61801, United States.

[4] Graduate School of Science and Technology, University of Tsukuba, 1-1-1 Tennodai, Tsukuba, Ibaraki, 305-8573, Japan.

[5] College of Materials Science and Engineering, Zhengzhou University, 450001, Zhengzhou, China.

[6] Department of Computer Science and Engineering, The Chinese University of Hong Kong, 999077, Hong Kong, China.

**Abstract**

Nanometric solid solution alloys are utilized in a broad range of fields, including catalysis, energy storage, medical application, and sensor technology. Unfortunately, the synthesis of these alloys becomes increasingly challenging as the disparity between the metal elements grows, due to differences in atomic sizes, melting points, and chemical affinities. This study utilized a data-driven approach incorporating sample balancing enhancement techniques and multilayer perceptron (MLP) algorithms to improve the model's ability to handle imbalanced data, significantly boosting the efficiency of experimental parameter optimization. Building on this enhanced data processing framework, we developed an entropy-engineered synthesis approach specifically designed to produce stable, nanometric copper and cobalt (CuCo) solid solution alloys. Under conditions of −0.425 V (vs. RHE), the CuCo alloy exhibited nearly 100% Faraday efficiency (FE) and a high ammonia production rate of 232.17 mg h$^{-1}$ mg$^{-1}$. Stability tests in a simulated industrial environment showed that the catalyst maintained over 80% FE and an ammonia production rate exceeding 170 mg h$^{-1}$ mg$^{-1}$ over a testing period of 120 hours, outperforming most reported catalysts. To delve deeper into the synergistic interaction mechanisms between Cu and Co, in situ Raman spectroscopy was utilized for real-time monitoring, and density functional theory (DFT) calculations further substantiated our findings. These results not only highlight the exceptional catalytic performance of the CuCo alloy but also reflect the effective electronic and energy interactions between the two metals.

# Introduction

Nanometric solid solution alloys are utilized in a broad range of fields due to their unique properties[1]. They are highly valued in catalysis for their ability to speed up chemical reactions[2,3], in energy storage[4] where they improve battery and fuel cell efficiencies, in medical applications where they are used for targeted drug delivery systems and antimicrobial coatings[5], and in sensor technology[6] where they enhance sensitivity and selectivity.

Nanometric solid solution alloys are crucial for various advanced applications[7], necessitating the exploration of a broader range of metallic combinations to maximize their potential[8]. However, significant challenges in synthesis arise due to substantial differences in atomic sizes, melting points, and chemical affinities among the constituent metals, which impede the development of these alloys. To overcome these obstacles, it is essential to employ more sophisticated synthesis techniques and innovative materials science concepts, ensuring that nanometric solid solution alloys achieve their superior performance in intended applications.

In the formation of nanoscale solid solution alloys, entropy plays a pivotal and critical role[9], serving not only as the main driving force behind alloy formation but also as a decisive factor in determining the thermodynamic stability of the alloy[7]. Particularly at the nanoscale, the increase in mixing entropy, resulting from the random mixing of different metal atoms, significantly enlarges the number of possible microstates of the system, thereby elevating the mixing entropy[10]. Nanometric solid solution alloys exhibit a pronounced surface entropy effect due to their high surface area-to-volume ratio, where the increased ratio of surface atoms to bulk atoms makes the contribution of surface entropy to the total entropy particularly significant[11]. The complex arrangement and distribution of surface atoms, influenced by surface activity and interface effects, become key variables in the formation and stability of solid solution phases at the nanoscale[12].

As the size of materials reduces to the nanometric level, the increase in surface energy and lattice distortion energy, caused by the rise in the number of surface atoms[13], typically drives the system towards a higher energy state[14]. However, the amplification of entropy at elevated temperatures, especially as temperature (T) increases, significantly enhances the contribution of entropy (S) to the free energy ($G = H - TS$), effectively reducing the system's free energy and facilitating the formation of solid solution phases[15]. The role of entropy is particularly pronounced during high-temperature synthesis processes, as elevated temperatures not only boost the growth of mixing entropy but also ease the redistribution of atoms within nanoparticles, promoting the formation of stable solid solution phases among different atoms[12].

In recent advancements, the fusion of neural networks with the realm of materials science and molecular investigation has sparked significant breakthroughs, demonstrating the immense potential of these computational tools[16]. The remarkable representational capabilities of neural networks have been pivotal in these achievements, inspiring their application to streamline the discovery of optimal experimental parameters and enhance the synthesis efficiency of nanoscale solid solution alloys[17].

Neural networks possess inherent qualities that make them exceptionally adept at navigating the complexities of experimental parameter selection for alloy synthesis[18]. They can adeptly discern the intricate and nonlinear relationships between various experimental conditions and the successful creation of nanoscale solid solution alloys, thereby offering precise predictions[19]. Moreover, neural networks are proficient at identifying patterns and correlations within data, allowing for a thorough analysis of how different experimental parameters influence outcomes[18]. Additionally, the flexibility and scalability of neural networks accommodate the continual expansion of experimental data, thereby progressively refining the accuracy of their predictions[20].

In this work, with objectives to lower experimental expenses and shorten development cycles, we utilized machine learning techniques to guide the selection of key experimental parameters, pioneering the development of an entropy-engineered synthesis approach for the preparation of stable nanometric CuCo solid solution alloys. This methodology, achieving solid solutions at the nanoscale for material combinations challenging to form at the macroscopic scale, not only offers new insights into the fundamental theories of material science but also lays the groundwork for the development of innovative catalysts. Moreover, the electrocatalytic performance of the CuCo alloy for the reduction of nitrate to ammonia (NO$_3$RR) was evaluated, revealing that at –0.425 V (vs. RHE), the CuCo alloy exhibited nearly 100% Faradaic efficiency (FE) and a yield of 232.17 mg h$^{-1}$ mg$^{-1}$ (including the mass of the carbon carrier). Stability tests conducted in a simulated industrial environment for up to 120 hours demonstrated that the catalyst not only maintained a FE exceeding 80% but also sustained an ammonia yield above 170 mg h$^{-1}$ mg$^{-1}$, surpassing the performance of the majority of catalysts reported in the literature. The synergistic interaction mechanisms between the active phases of copper and cobalt were studied using in situ Raman spectroscopy with density functional theory (DFT) calculations providing further validation of our findings.

## 2 Experimental Section

### 2.1 Materials

Anhydrous cobalt chloride (CoCl$_2$, 99.7%)and anhydrous copper chloride (CuCl$_2$, 98%) were acquired from Shanghai Macklin Biochemical Co., Ltd. N,N-dimethylformamide (DMF, AR) was supplied by Sinopharm Chemical Reagent Co., Ltd. Polyacrylonitrile (PAN, Mw=1.49×10$^5$, copolymerized with 10wt% acrylate) was obtained from Sinopec Shanghai Petrochemical Co., Ltd. Potassium hydroxide (KOH, AR≥90%) was purchased from Shanghai Macklin Biochemical Co., Ltd., and potassium nitrate (KNO$_3$, AR) was obtained from Sinopharm Chemical Reagent Co., Ltd. Isopropanol (AR, ≥99.7%) was procured from Shanghai Macklin Biochemical Co., Ltd. Nafion 117 solution (5wt%) was supplied by Shanghai Aladdin Biochemical Technology Co., Ltd. Ammonium chloride (NH$_4$Cl, AR) and potassium nitrite (KNO$_2$, GR ≥97%) were acquired from Shanghai Aladdin Biochemical Technology Co., Ltd. N-(1-Naphthyl)ethylenediamine dihydrochloride (AR, ≥97%), sulfanilamide (AR≥99.8%), phosphoric acid (H$_3$PO$_4$, AR≥85.0%), hydrochloric acid (HCl, AR), salicylic acid (AR), and sodium citrate dihydrate (Na$_3$C$_6$H$_5$O$_7$·2H$_2$O, 99.0%) were all purchased from Sinopharm Chemical Reagent Co., Ltd. Sodium hydroxide (NaOH, AR,≥96%) was obtained from Shanghai Titan Technology Co., Ltd. Sodium nitroprusside (AR, 99.0%) was procured from Shanghai Macklin Biochemical Co., Ltd. Sodium hypochlorite (NaClO, AR) was supplied by Shanghai TCL Chemical Industrial Development Co., Ltd. N$_{15}$ potassium nitrate (K$^{15}$NO$_3$, 99atom%;>99.0%) was obtained from Shanghai Aladdin Biochemical Technology Co., Ltd. $N_{14}$ potassium nitrate ($K^{14}NO_3$, 99atom%;>99.0%) was obtained from Shanghai Aladdin Biochemical Technology Co., Ltd. Maleic acid ($C_4H_4O_4$, 99%) was obtained from Shanghai Aladdin Biochemical Technology Co., Ltd.

## 2.2 Preparation of nanometric solid solution CuCo alloys

### 2.2.1 Synthesis of the Cu/Co single-metal / CuCo alloy catalysts

To prepare a uniform and transparent CuCo salt/PAN (polyacrylonitrile) spinning solution, 1.5 mM of $CoCl_2$, 1.5 mM of $CuCl_2$, and 3g of PAN are dissolved in 30g of DMF (N,N-Dimethylformamide). This solution is then stirred magnetically at room temperature for 12 hours. Following this, the solution is loaded into a syringe equipped with a stainless-steel needle for electrospinning, leading to the formation of CuCo/PAN nanofiber membranes. The electrospinning process is optimized by setting the anode voltage to 20 kV, the flow rate to 0.3 mL/h, and the distance between the needle and the collector to 15 cm, within an environment maintained at 45°C and 20% humidity. Subsequently, the nanofiber membranes are cut into 10×10 $cm^2$ rectangular shapes and placed in an oven. The temperature in the oven is gradually increased from 60°C at a rate of 10°C/h until it reaches 240°C for pre-oxidation, and the membranes are held at this temperature for 3 hours to complete the pre-oxidation treatment. Following pre-oxidation, the membranes are then heated in an argon atmosphere from room temperature to 1000°C at a rate of 2°C/min and maintained at 1000°C for 5 hours. To finalize the synthesis, the samples are rapidly cooled by being placed in a liquid nitrogen-filled environment directly from the high temperature of 1000°C, resulting in the creation of CuCo alloy nanoparticles embedded in nitrogen-doped carbon fibers. The synthesis of single-metal catalysts based on copper or cobalt shares similarities with the preparation of CuCo alloy catalysts.

### 2.2.2 Synthesis of the CuCo dual phase catalysts

Following the procedure used for CuCo alloy catalysts, the furnace is shut off after a 5-hour period at 1000 degrees Celsius, enabling a slow return to ambient temperature.

### 2.3 Characterization

The morphology of the as-prepared catalysts was investigated by scanning electron microscopy (SEM) and transmission electron microscopy (TEM). SEM was conducted on Hitachi S-4800 FE-SEM with an acceleration voltage of 5 kV and a current of 10 μA. TEM, high-angle annular dark-field scanning transmission electron microscopy (HAADF-STEM) and energy dispersive X-ray analysis were performed on a Titan Cubed Themis G2300 equipped with Cs corrector and energy-dispersive X-ray spectrometer at 200 kV. The X-ray diffraction (XRD) patterns of the catalysts were collected on Bruker D8 Advance X-ray diffractometer. X-ray photoelectron spectroscopy (XPS) was performed on Thermo Scientific K$_\alpha$ system with the C 1s peak (284.8 eV) as reference. The Raman spectra were recorded on Renishaw inVia confocal microscope using excitation wavelength of 785 nm. The Raman measurements were carried out over a range of 4000 to 100 cm$^{-1}$.

### 2.4 Machine Learning Algorithms

Our dataset is composed of experimental cases from our laboratory's past research. These data points have been systematically recorded and are now being utilized to inform our predictive models. When dealing with complex tasks such as alloy synthesis prediction, the issue of class imbalance can significantly impede the model's ability to learn effectively. Class imbalance occurs when the dataset contains significantly more instances of one class compared to others, leading to a biased model that favors the majority class. This is a common challenge in datasets where successful synthesis outcomes are rare compared to unsuccessful ones.

For balanced sample augmentation, we first normalize each feature within every sample and then utilize the SMOTE and Tomek links. This normalization ensures that all feature values are scaled similarly, preventing any individual feature from disproportionately influencing model training due to its larger numeric range. Specifically, we compute the mean and standard deviation for each feature dimension and adjust each feature by subtracting its mean and dividing by its standard deviation. This standardizes the features to a normal distribution with a mean of 0 and a variance of 1.

For multi-layer perceptron (MLP), each sample's input is represented as a vector $x = (x_1, x_2, ..., x_n)$, consisting of $N$ features. The final output $\hat{y}$, representing the predicted probability of the sample relating to alloy synthesis success, is computed as $\hat{y} = MLP(x_1, x_2, ..., x_n)$. Here, the output $\hat{y}$ is a scalar value that indicates the model's predicted probability for the sample belonging to a certain class, typically representing the likelihood of successful alloy synthesis. We employed the Binary Cross-Entropy (BCE) loss function to minimize the difference between the model's predicted output $\hat{y}$ and the true labels $y$ as,

$$\text{BCELoss} = -\frac{1}{M}\sum_{i=1}^{M}[y_i\log(\hat{y}_i) + (1-y_i)\log(1-\hat{y}_i)]$$

where $M$ denotes the number of samples, $y_i$ denotes the true label of the $i$-th sample (zero or one), and $\hat{y}$ denotes the predicted probability by the model for the $i$-th sample.

For training, we employ the Adam optimizer with PyTorch on a single GTX4090Ti GPU, using a learning rate of 0.001, $\beta_1 = 0.900$, $\beta_2 = 0.999$, and a batch size of 16. Training is conducted for 50 epochs with an early stopping strategy implemented to prevent overfitting.

For validation, to ensure the model's generalization and robustness in the context of alloy synthesis prediction, we utilize Stratified K-Fold Cross Validation. The method ensures that each fold is representative of the overall dataset by maintaining the same proportion of successful and unsuccessful synthesis outcomes. Specifically, we divide the dataset into 5 subsets (folds), train the model on 4 folds, and validate the model on the remaining fold, repeating the process 5 times with different validation folds. This helps mitigate the issue of having a small dataset by maximizing the use of available data, providing a more accurate and unbiased assessment of the model's predictive capabilities for alloy synthesis success.

## 2.5 Electrochemical NO$_3$RR Measurements

The electrochemical evaluation was conducted in an H-type cell equipped with a three-electrode setup, utilizing a CHI 760E electrochemical workstation (manufactured by Shanghai Chenhua). A Nafion 117 membrane facilitated the separation within the H-type cell. Notably, the system was outfitted with a magnetic stirrer operating at a speed of 1000 rpm. The preparation method for the working electrode is as follows: initially, 3 mg of catalyst powder is dispersed in a solvent mixture comprising 500 μl isopropanol, 470 μl ultrapure water, and 30 μl of 5% Nafion solution, followed by ultrasonication for a minimum of one hour to ensure a uniform ink; subsequently, 5 μl of this ink is applied onto a glassy carbon electrode with an area of 0.196 cm², achieving a precise catalyst loading of 0.0765 mg/cm². The glassy carbon electrode laden with the catalyst serves as the working electrode, with a platinum wire and an Hg/HgO electrode filled with 1M KOH solution acting as the counter and reference electrodes, respectively. Prior to testing, all catalysts undergo electroreduction at –0.2 V versus RHE for 600 seconds in 1 M KOH solution to remove surface oxides. The electrolyte employed is a 1 M KOH solution containing varying concentrations of $KNO_3$ (volume of 30 ml). Before the commencement of experiments, the electrolyte is saturated with argon gas for 10 minutes to eliminate dissolved $O_2$ and $N_2$. Electrochemical Linear Sweep Voltammetry (LSV) curves were obtained in a single cell setup. Within a 1 M KOH solution, cyclic voltammetry curves for the determination of the electrochemical double-layer capacitance ($C_{dl}$) were measured at scan rates of 20, 40, 60, 80, and 100 mV/s under a potential window where no Faradaic processes occur. The formula for converting electrode potential to the Reversible Hydrogen Electrode (RHE) is: $E_{RHE} = E_{Hg/HgO} + 0.197 \text{ V} + 0.0591 \times \text{pH}$. It's noteworthy that in a 1 M KOH solution, the volatilization of $NH_3$ during a 2-hour electrolysis process is almost negligible. For extended electrolysis durations of up to 12 hours, a gas scrubber is employed to ensure experimental accuracy. Throughout the entire electrochemical testing phase, a continuous supply of argon gas maintains electrolyte saturation, thereby eliminating potential interference from $O_2$ and $N_2$.

## 2.6 Determination of ion concentrations

### 2.6.1 $NH_4^+$ Quantification[21]

A sample of the electrolyte is taken from the reaction cell and diluted to 2 ml. To this, 0.4 ml of a 1 M NaOH solution containing sodium citrate and salicylic acid (stored at 4°C) is added, followed by 200 μl of freshly prepared 0.05 M NaClO (AR). The mixture is then thoroughly shaken. After this, 0.2 ml of a 1 wt.% sodium nitroprusside solution (also stored at 4°C) is introduced to catalyse the colorimetric reaction. The mixture is left to stand at room temperature for 2 hours before its absorbance at 655 nm is measured using a UV-vis spectrophotometer to determine $NH_3$ concentration. A calibration curve, created using a standard $(NH_4)_2SO_4$ solution in 1 M KOH, facilitates accurate $NH_3$ quantification.

### 2.6.2 $NO_2^-$ Quantification[22]

A portion of the electrolyte is diluted to 2 ml and treated with 0.2 g of N-(1-naphthyl) ethylenediamine dihydrochloride, 4.0 g of sulfanilamide, and a mixture of 10 ml of 85 wt.% phosphoric acid in 50 ml of deionized water. Initially, 1 ml of 1 M $H_2SO_4$ is added to 5 ml of diluted electrolyte, followed by 0.1 ml of the color reagent, ensuring a homogenous mixture.

After 20 minutes at room temperature, the solution's UV-vis absorbance at 540 nm is recorded. The $NO_2^-$ concentration is determined using a calibration curve based on $NaNO_2$ solutions. Notably, while compounds such as $N_2H_4$ and $NH_2OH$ may form during nitrate electroreduction, their reactivity in alkaline conditions renders their concentrations minimal and typically only detectable at intermediate stages. The focus thus remains on $NH_3$ and $NO_2^-$ yields.

## 2.7 Efficiency calculation

Definition of FE: The Faradaic Efficiency (FE) quantifies the proportion of electrical charge used specifically for generating a desired product, like $NH_3$, relative to the total charge ($Q$) transferred across the electrodes during electrolysis.

Calculation of FE for $NH_3$ ($FE_{NH3}$):

Given that the production of a single $NH_3$ molecule requires the consumption of eight electrons, the FE for $NH_3$ can be determined using the following equation:

$$FE_{NH3} = \frac{8 \times F \times C_{NH3} \times V}{Q} \qquad (1)$$

Where $F$ denotes the Faraday constant. $C_{NH3}$ is the molar concentration of detected $NH_3$. $V$ represents the volume of the electrolytes. $Q$ is the total charge passing through the electrodes.

Calculation of Ammonia mass Yield ($Y_{massNH3}$): The yield of $NH_3$ can be calculated as:

$$Y_{massNH3} = \frac{C_{NH3} \times V}{A \times t} \qquad (2)$$

Where $A$ represents the mass of the catalyst. $t$ is the duration of the reaction. Calculation of

Ammonia area Yield ($Y_{areaNH3}$): The yield of $NH_3$ can be calculated as:

$$Y_{areaNH3} = \frac{C_{NH3} \times V}{B \times t} \quad (3)$$

Where *B* represents the area of the electrode.

### 2.8 $^{15}NO_3^-$ Isotope labelling experiments and $^{14}NH_3$ quantification by $^1$H NMR[23]

In isotopic labeling experiments for nitrate reduction, K$^{15}$NO$_3$ (99 atom%; >99.0%) was used as the nitrogen source to trace the formation of ammonia. A 1 M KOH solution served as the electrolyte, with 0.1 M K$^{15}$NO$_3$ added to the cathode compartment as the reactant. After electroreduction, the electrolyte containing the resultant $^{15}NH_4^+$ was collected, and its pH was adjusted to 3.5 using 4 M H$_2$SO$_4$. Subsequently, quantification was performed via $^1$H NMR (600 MHz), employing maleic acid as an external standard. Similarly, the presence of $^{14}NH_4^+$ was assessed using the same approach when K$^{14}$NO$_3$ was utilized as the nitrogen source.

### 2.9 In situ Raman spectroscopy[21]

Raman spectroscopy analyses were conducted using a Lab-RAM Raman microscopy system (Horiba Jobin Yvon, HR550), featuring a 532 nm laser for excitation, a water immersion objective (Olympus LUMFL, 60× with a numerical aperture of 1.10), an 1800 grooves/mm grating monochromator, and a Synapse CCD detector. Each recorded spectrum represents the average of five spectra acquired consecutively, with each having a collection time of 50 seconds. For in situ Raman experiments, a three-electrode electrochemical cell was utilized, employing

Pt wires and Ag/AgCl (3 M KCl) as counter and reference electrodes, respectively. To safeguard the objective lens from the corrosive nature of 0.1 M KOH electrolyte, a less concentrated 0.01 M KOH solution (pH = 12) was used instead. $K_2SO_4$ (⩾99.0%) was incorporated to maintain adequate ionic conductivity by keeping the total $K^+$ concentration at 0.1 M, and to provide $SO_4^{2-}$ ions as an external Raman standard. For experiments involving 0.01 M $KNO_3$, the supporting electrolytes consisted of 0.01 M KOH and 0.04 M $K_2SO_4$. In setups devoid of $KNO_3$, the electrolyte mixture was 0.01 M KOH and 0.045 M $K_2SO_4$.

**2.10 DFT calculations**:

Density Functional Theory (DFT) calculations were performed using the Vienna *ab initio* Simulation Package (VASP)[24]. We employed the generalized gradient approximation (GGA) with the Perdew-Burke-Ernzerhof (PBE) formulation for the exchange-correlation functional. Electron-ion interactions were modeled using the projector augmented wave (PAW) method [25]. The calculations utilized a plane wave basis with a 450 eV cutoff [26]. Convergence of forces was assured at 0.02 eV/Å [13] with the electronic convergence of $10^{-6}$ eV [27]. To minimize periodic image interactions, a 20 Å vacuum layer was incorporated [28,29]. Dispersion corrections were applied using Grimme's D3 method with Becke-Jonson damping to accurately account for van der Waals forces. The computational approach was validated through stringent convergence tests and comparisons with experimental benchmarks.

The Gibbs free energy of $\Delta G$ is shown below [30]:

$$\Delta G^* = \Delta E + \Delta E_{ZPE} - T\Delta S + \Delta U_{0\to 300K}$$

where $\Delta E$, $\Delta E_{ZPE}$, $\Delta S$, $\Delta U_{0\to 300K}$, and $T$ represent the calculated reaction energy, the zero-point energy correction, the entropy differences, the thermal correction to energy under 300 K without ZPE, and the temperature (300 K), respectively.

## 3 Results and Discussion

### 3.1 Design Principle and Structural Characterizations

The formation of solid solutions typically depends on several key conditions: compatibility of crystal structures, minor differences in atomic sizes, similarity in electronegativity among elements, and the effect of valency[31]. These factors work together to determine whether alloy components can form continuous solid solutions with high solubility. At the nanoscale, previous research findings and our laboratory experience provide crucial guidance for understanding and facilitating the formation of solid solution alloys. However, when exploring a broader range of solid solution alloy possibilities, the synthesis experiments for materials become particularly challenging due to the long experimental cycles, high costs, and the high price of trial and error.

In this work, we introduce a data-driven algorithm designed to predict the success of alloy synthesis. This algorithm serves as a tool for identifying optimal experimental parameters, thereby enhancing the efficiency of practical experimentation, as illustrated in **Figure 1**a.

Specifically, we introduce a balanced sample augmentation method to increase the number of training samples and balance the two types of samples. We combine two data augmentation techniques, namely SMOTE (Synthetic Minority Over-sampling Technique) and Tomek links. SMOTE is utilized to generate synthetic samples for the minority class, achieving sample balance, as shown in **Figure 1**b. Tomek links, on the other hand, are employed to clean the data by identifying and removing majority class samples near the decision boundary of the minority class. As a result, we effectively increase the number of samples and balance the dataset, enriching the sample diversity and facilitating subsequent neural network learning.

Based on the augmented data, we employ a multi-layer perceptron (MLP) to predict the success probability of alloy synthesis, as illustrated in **Figure 1**c. The MLP is a feedforward neural network composed of multiple fully connected layers, with each neuron connected to all neurons in the previous layer. It includes input layers, several hidden layers, and an output layer. In each hidden layer, we use common activation functions (i.e., ReLU) to introduce nonlinearity, enhancing the network's representation ability. The output layer utilizes a sigmoid activation function (**Figure 1**d) to constrain outputs between 0 and 1, representing the probabilities of different classes. We use the cross-entropy loss function to measure the difference between model outputs and true labels, and update network parameters through a backpropagation algorithm to minimize the loss function.

We conduct training on a mixed dataset comprising 'CuAu', 'CuAg', 'CuCr', and 'CuMn' alloy synthesis, with approximately 100 augmented training samples. As depicted in **Figure 1**e, we select representative experimental parameters as inputs and feed them into the MLP for training. We validate the model's classification accuracy on CuCo alloy synthesis. Our AUC metric reaches 0.97, demonstrating the effectiveness of our algorithm and its effective guidance in experimental efficiency.

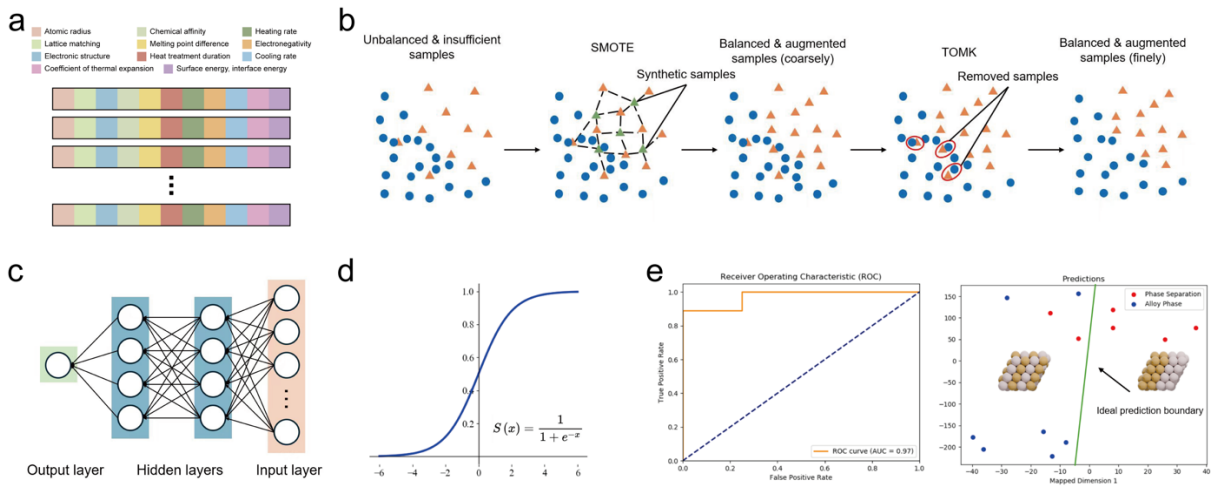

Figure 1. Machine Learning-Guided Material Synthesis (a) Several key conditions for the formation of alloy phase. (b) Balanced sample augmented algorithm. (c) Multi-layer perceptron. (d) Sigmoid function that maps latent features to a probability distribution between 0 and 1. (e) left: the orange ROC curve plots the true positive rate against the false positive rate at different classification thresholds, where RUC is 0.97 (higher is better); right: prediction results of our model.

3.2 Synthesis and characterizations of CuCo Alloy

**Figure 2**a illustrates the entropy-engineered synthesis of stable ultrafine nanometric solid solution alloys, beginning with the production of nanofiber membranes from a mixture of polyacrylonitrile (PAN) and two types of metal salts using electrospinning technology[32]. Initially, the pre-oxidation phase is dedicated to the removal of residual solvents such as DMF, a crucial step for purifying the structure[33]. During this phase, the PAN polymer chains undergo oxidation, incorporating oxygen atoms to form cross-linked structures, a process referred to as stabilization. This not only strengthens the polymer matrix but also enhances its resistance to decomposition in subsequent high-temperature carbonization processes. Concurrently, the added metal salts are converted into metal oxides. This process is visually signified by the transition of the fiber membranes' color from their original pale yellow to a deep brown, indicating the formation of conjugated structures and preliminary carbides. At the microscopic level, these changes result in a denser and more ordered fiber membrane structure, improving the electrical conductivity and structural integrity of the final carbon materials.

When two pure elements, let's call them A and B, are combined to form an alloy, the resulting material's structure and properties depend significantly on the interactions between A and B at the atomic level[34]. These interactions are largely governed by the relative free energies of the various possible combinations and arrangements of A and B atoms[35,36]. Depending on these energies, the alloy can form a solid solution, one or more intermetallic phases, or a combination of both:

$$\Delta G_{mix} = \Delta H_{mix} - T\Delta S_{mix} \qquad \text{Eq. (1) solid solution mixing}$$

$$\Delta G_{inter} = \Delta H_{inter} - T\Delta S_{inter} \qquad \text{Eq. (2) intermetallic phases } A_iB_j$$

where $\Delta G_{mix}$, $\Delta H_{mix}$, and $\Delta S_{mix}$ are the change in Gibbs free energy, enthalpy, and entropy for the formation of a solid solution, respectively; $\Delta G_{inter}$, $\Delta H_{inter}$, and $\Delta S_{inter}$ are the corresponding values for the formation of an intermetallic compound with stoichiometry (that is, $A_iB_j$, where *i, j* = 1, 2, 3…). T is the temperature[37].

In deciding whether an alloy will form a solid solution, intermetallic phases, or a mixture of both, one must compare $\Delta G_{mix}$ and $\Delta G_{inter}$ at thermodynamic equilibrium. The phase (or phases) with the lower Gibbs free energy at a given temperature and composition will be more stable and is thus more likely to form[38].

Indeed, the formation of intermetallic compounds within an alloy consisting of elements A and B does not always lead to the complete transformation of the alloy into a single phase of an intermetallic compound $(A_iB_j)$[39]. Instead, intermetallic can precipitate within a solid solution that is rich in either A or B, in which the Gibbs free energy change involves both the free energy of mixing to form the terminal solid solution and the free energy of formation of the intermetallic phase[40]. If the combined process of mixing and intermetallic formation is energetically favorable, and the alloy will tend to form a mixture of a terminal solid solution with precipitates of the intermetallic phase. The overall stability and phase distribution within the alloy will depend on these Gibbs free energies[41].

During the heating process, polymer nanofibers undergo carbonization, while metal salts experience carbothermal reduction[42]. The thermal decomposition of polymer chains releases small molecules such as water, carbon monoxide, and carbon dioxide, increasing the system's disorder and entropy[43]. As carbonization progresses, the newly formed carbonaceous structure transitions into a more ordered graphitic state, potentially leading to a decrease in system entropy. Simultaneously, metal ions gain electrons and are released from their orderly lattice state to become free metal atoms[44]. Their disorderly dispersion on the carbon fiber structure further increases local entropy. Throughout the carbothermal reduction process, metal atoms aggregate into nanoparticles via the Ostwald ripening mechanism, demonstrating the system's dynamic entropy balance[45]. Initially, the reduction of metal ions to free metal atoms and their dispersion introduces a greater number of possible microstates, resulting in an increase in entropy[46,47]. However, as these metal atoms gradually aggregate, forming larger nanoparticles and reducing surface energy, local entropy seemingly decreases[48]. Yet, this transition towards a more stable particulate form, from a macroscopic perspective, still aligns with the system's pursuit of entropy increase to achieve thermodynamic stability[49]. After the complete reduction of metal salts, the continued heating and injection of external energy elevate the system's overall energy state, endowing metal atoms with increased kinetic energy[50]. This energy input and rearrangement of metal atoms not only lead to an increase in system entropy but also facilitate potential interactions and alloying between metal atoms, further influencing particle morphology and size[51,52]. In this high-temperature state, according to the principle of entropy increase, metals tend toward an ideal alloy state, undoubtedly also a high entropy state[53]. Rapid cooling techniques maintain the system at this high entropy level by limiting the diffusivity of metal atoms, effectively "freezing" their high-temperature disordered distribution[54]. This process locks the system into a non-equilibrium but high entropy state rapidly, preventing a further decrease in entropy, i.e., a transition to a more ordered state, thereby inhibiting significant phase separation. Conversely, slow cooling allows metal atoms ample time to diffuse and rearrange into lower energy, more stable configurations[54]. In this process, Co and Cu atoms tend to separate based on their solubility and thermodynamic driving forces, forming a phase-separated structure. This phase separation process illustrates the system's transition from disorder to order, accompanied by a decrease in entropy, ultimately achieving thermodynamic stability. The theory of entropy change provides a logical framework for understanding the different microstructural outcomes resulting from rapid and slow cooling: rapid cooling preserves a high entropy non-equilibrium state to avoid phase separation, while slow cooling promotes entropy reduction and phase separation, leading to thermodynamic equilibrium. This analysis reflects the core of the second law of thermodynamics, that systems always evolve towards the state of lowest energy and highest entropy, with the cooling rate critically influencing the specific path to this state[55]. As shown in the Figure S1, the carbon nanofibers loaded with CuCo alloy have a diameter of approximately 100–200 nm. **Figure 2**b shows the X-ray diffraction (XRD) patterns of Cu, Co, and CuCo. The diffraction peaks for Cu are observed at $2\theta$ values of 43.3°, 50.4°, and 74.1°, which match the (111), (200), and (220) planes, respectively, and correspond to PDF#04-013-4520. The Co pattern displays peaks at 44.2°, 51.5°, and 75.8°, corresponding to PDF#04-015-0419. The CuCo alloy exhibits distinct peaks at 43.9° and 50.7°, corresponding to the (111) and (200) planes of the alloy and matching PDF#04-020-2827. These results confirm the successful synthesis of the solid solution CuCo alloy, as the observed peaks align precisely with the reference patterns. **Figure 2**c–e present transmission electron microscopy (TEM) images of the synthesized the solid solution CuCo alloy, demonstrating its morphology and crystal structure. **Figure 2** c shows the overall morphology of the CuCo alloy nanoparticles, which appear to be well-dispersed with sizes in the nanometer range. **Figure S3** shows a normal distribution of particle sizes, concentrated between 10 and 16 nm, with the highest frequency of particles at 12 nm. As shown in **Figure 2**d–e, the CuCo alloy is supported on carbon nanofibers. By combining High-Resolution Transmission Electron Microscopy (HRTEM) and Selected Area Electron Diffraction (SAED), it can be determined that the structure in the image is single crystalline. Therefore, it can be inferred that Co and Cu exist in the form of an alloy. The measured lattice fringe spacing is 0.216 nm corresponds to the (111) planes of the CuCo alloy, consistent with the interplanar spacing of the CuCo alloy reported in the literature[56]. Complementary energy-dispersive X-ray (EDX) mapping reveals a consistent distribution of both Cu and Co elements, cementing the formation of the CuCo solid solution alloy (**Figure 2**f).

Raman spectroscopy analysis (**Figure S4**) reveals that the $I_D/I_G$ ratios for Cu single metal, CuCo alloy, and Co single metal samples are 1.166, 1.169 and 1.173, respectively. These values indicate a variation in defect density within the carbon structure associated with different metal types. The lowest $I_D/I_G$ ratio for the Cu single metal suggests a higher degree of graphitization and fewer structural defects. In contrast, the moderate $I_D/I_G$ ratio for the CuCo alloy may reflect the contribution of alloying to the increase of defects in the carbon structure. The highest $I_D/I_G$ ratio for the Co single metal indicates a significant increase in defects within its carbon structure.

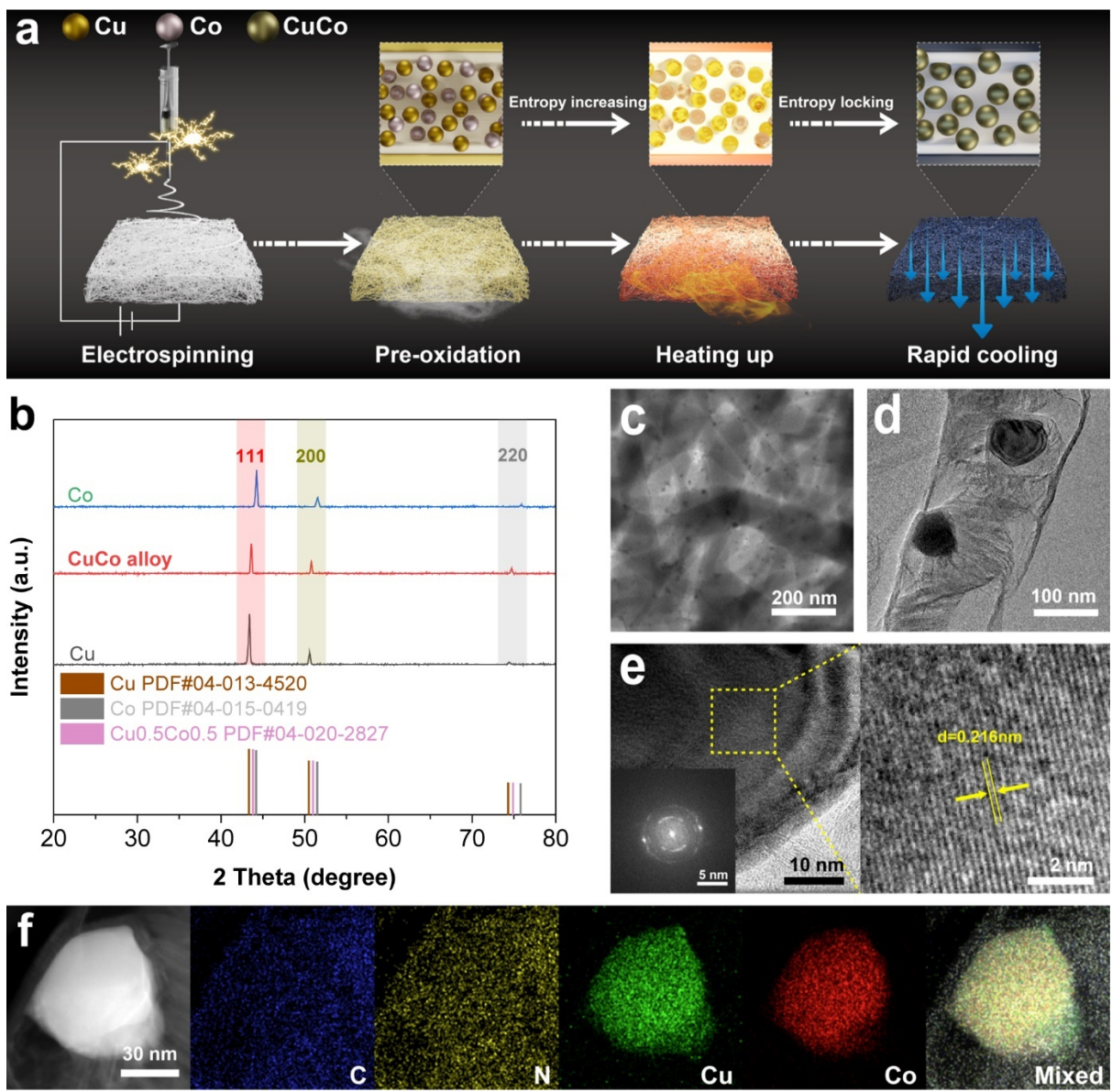

Figure 2. Preparation Processes and Characterization (a) Scheme of Entropy-Engineered Synthesis of Stable Ultrafine Nanoalloys. (b) XRD patterns of the CuCo alloy and its comparative samples. (c) TEM image of CuCo Alloy@CNFs (d) TEM image of CuCo Alloy@CNFs (e) AC-HAADF-STEM image of CuCo Alloy@CNFs. (f) AC-STEM images of CuCo Alloy@CNFs, and the STEM-EDS elemental mapping images. The insets in e show the corresponding FFT patterns.

3.3 Catalytic performance of CuCo Alloy for the NO$_3$RR

In pursuit of a comprehensive understanding of the efficacy of CuCo alloy electrocatalyst in catalyzing $NO_3RR$, this study embarked on a detailed analysis by comparing the performance of CuCo alloy with Cu single metal and Co single metal. Linear sweep voltammetry (LSV) was employed to evaluate the activity of the catalysts in 1 M KOH solution (pH 14) and in 1 M KOH solution supplemented with varying fixed concentrations of $NO_3^-$. It is important to note that, unless specifically stated otherwise, all potential values were calibrated against the reversible hydrogen electrode (RHE).

As illustrated in **Figure 3**a, under conditions of increasing applied potential, the same catalyst demonstrated significantly higher current densities in the presence of $NO_3^-$ compared to environments devoid of $NO_3^-$. This observation strongly suggests that in 1 M KOH solution without $NO_3^-$, the catalyst predominantly facilitates the hydrogen evolution reaction (HER). Conversely, the introduction of $NO_3^-$ alters the pathway of the catalytic reaction to involve the nitrate reduction reaction. Under both conditions (with and without $NO_3^-$), CuCo alloy exhibited significantly higher current densities compared to Cu single metal and Co single metal, indicating its superior catalytic activity. Further investigation into the catalytic activity of the CuCo alloy, as shown in **Figure 3**b, was conducted through LSV tests in 1 M KOH solution with varying concentrations of $NO_3^-$ added. A notable increase in current density was observed for the CuCo alloy, indicating a high sensitivity towards $NO_3^-$, which suggests a strong adsorption capacity for $NO_3^-$. With increasing concentrations of $NO_3^-$, a more pronounced enhancement in current density was observed, signifying that the activity of $NO_3RR$ intensifies with higher $NO_3^-$ concentrations. Employing an electrolyte solution of 1 M KOH + 0.1 M $KNO_3$, CuCo alloy was subjected to current-time (It) curve testing across various potentials (−0.15 to −0.5V) over a duration of 2 hours, during which the faradaic efficiency (FE) and yield of ammonia ($NH_3$) at different potentials were calculated. As depicted in **Figure 3**c, CuCo alloy achieved nearly 100% FE and a yield of 232.17 mg h$^{-1}$ mg$^{-1}$ (including the mass of the carbon carrier in the catalyst mass) at −0.425V. With an increase in voltage, the current density in the It curve increased, and beyond −0.45V, significant fluctuations in the It curve were observed (**Figure S5**), attributed to the competitive HER, leading to the release of hydrogen gas at the electrode surface, consistent with the observed decrease in NH$_3$ FE. This phenomenon validates the decrease in NH$_3$ FE, reflecting the negative impact of HER competition on NH$_3$ synthesis efficiency at higher voltages.

NO$_3$RR is a complex process involving multi-electron transfers and numerous competing reactions[57]. This reaction follows a coherent pathway where nitrite (NO$_2^-$) emerges as a stable intermediate. The efficient conversion of NO$_3^-$ to NH$_3$ necessitates accelerated reactions from NO$_3^-$ to NO$_2^-$ and subsequently from NO$_2^-$ to NH$_3$. At lower potentials, the reaction tends towards the formation of NO$_2^-$, hindering the complete reduction of NO$_3^-$ to NH$_3$. Conversely, as the potential increases, the competitive HER becomes predominant, thereby reducing energy utilization efficiency.

By assessing the product selectivity of Cu, Co, and CuCo alloy at different potentials using UV-Vis spectrophotometry (**Figure S5**), we observed significant differences in the FE for NH$_3$ and NO$_2^-$ (**Figure 3**d-f and **S6-7**). Under lower overpotential conditions, Cu exhibited a stronger propensity to facilitate the formation of NO$_2^-$ compared to Co, indicating a greater adsorption and conversion capacity of Cu for NO$_3^-$ and an effective suppression of HER. In contrast, the It curves of Co at the same potentials were more fluctuant, reflecting a more pronounced HER activity and proton (H$^+$) evolution capability compared to Cu. CuCo alloy catalyst combines the advantages of Cu and Co, displaying a notable synergistic effect. In a high-entropy state, the elemental mixing between Cu and Co sites is significantly high, maximizing the synergy. Notably, the cooperative interaction between Cu and Co sites leads to higher $NH_3$ FE at elevated potentials, implying a higher yield. This synergistic effect not only optimizes the structural characteristics of the catalyst but also significantly enhances its catalytic performance, as evidenced by the exceptionally high $NH_3$ faradaic efficiency ($NH_3$ FE) and the remarkably low $NO_2^-$ formation efficiency, indicating the maximization of Cu and Co site cooperation.

To unequivocally determine the source of synthesized $NH_3$ as being from $NO_3^-$, isotope labeling studies were conducted using both $^{14}NO_3^-$ and $^{15}NO_3^-$ isotopes. The identification and quantification of the products were carried out through proton nuclear magnetic resonance ($^1H$ NMR) spectroscopy. When employing $^{14}NO_3^-$, the $^1H$ resonance displayed a unique tripartite splitting, clearly indicating the formation of $^{14}NH_4^+$. In contrast, the use of $^{15}NO_3^-$ resulted in the resonance splitting into two distinct signals, affirmatively confirming the generation of $^{15}NH_4^+$ (**Figure 3**g). Furthermore, to ensure the integrity and accuracy of the data collected, the indophenol blue method (IBS) was primarily utilized for the detection of $NH_3$.

To evaluate the stability of the catalyst and its potential for industrial applications, this study conducted It curve tests at a set potential of −0.425V in a solution containing 1 M KOH and 0.1 M $KNO_3$, lasting for 12 hours across 10 cycles. As depicted in **Figure 3**h, the It curves exhibited a consistent trend with the progression of reaction time, specifically a gradual decrease in current density as the $NO_3RR$ proceeded. This phenomenon is primarily attributed to the substantial consumption of $NO_3^-$ (the sole nitrogen source) in the reaction. The lack of timely replenishment of the nitrogen source led to a decrease in reactant concentration, thereby reducing the current density. **Figure 3**i illustrates that, after the extended 12-hour It curve testing, the catalyst not only maintained a FE exceeding 80% but also sustained an ammonia yield above 170 mg h$^{-1}$ mg$^{-1}$, without any observable signs of catalyst deactivation. This suggests the catalyst's potential for long-term stability. The catalyst subjected to a 120-hour It curve test is further characterized to evaluate its microstructure. SEM images confirm the preservation of its fibrous morphology (**Figure S8**). TEM images further reveal uniform distribution of catalyst particles on carbon fibers (**Figure S9**). XRD analysis (**Figure S10**) both confirm the alloy structure remained stable throughout the extended reaction period, with no phase separation observed, further validating the exceptional stability of the catalyst.

Given the significant impact of $NO_3^-$ concentration on the variation of current density, this study utilized 0.1 M $NO_3^-$ as the benchmark to assess and compare the performance of our catalyst against those reported in the literature in terms of area yield and FE (**Figure S11**). The results distinctly demonstrate the significant potential of the CuCo alloy catalyst, which showcased exceptional FE and high yield, highlighting its promising application prospects in the field of electrochemical catalysis.

Therefore, the design of the CuCo alloy is not solely based on enhancing the catalytic performance of a single metal but also leverages the interactions between metals to achieve precise control over the reaction pathway, significantly improve catalytic efficiency, and maintain robust stability under high currents. This offers an effective strategy for efficient electrocatalytic $NO_3RR$.

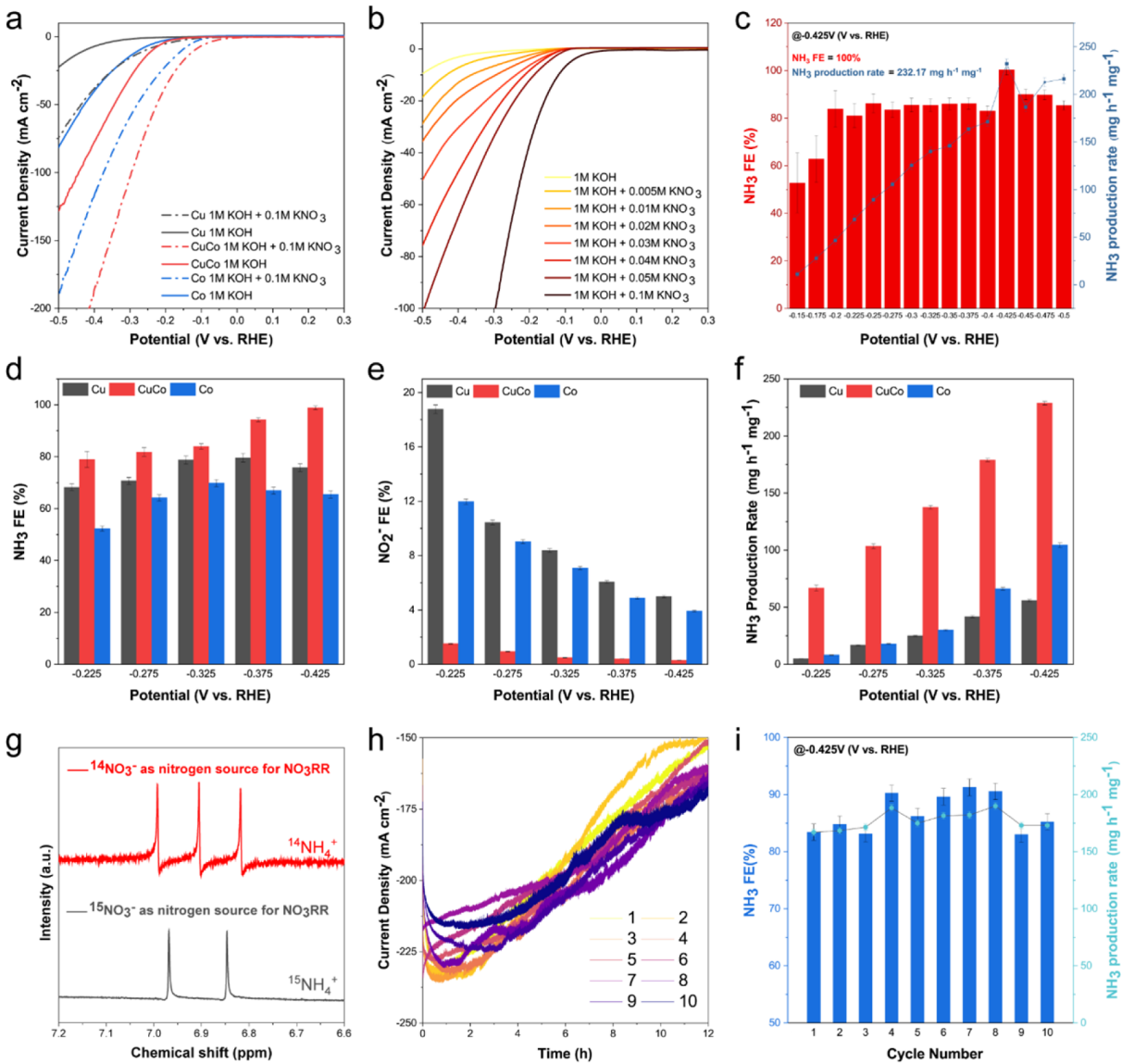

Figure 3. Electrocatalytic performance of CuCo alloy for NO₃RR (a) Linear sweep voltammograms of Cu single metal, Co single metal, and CuCo alloy in both 1 M KOH electrolyte and 1 M KOH with 0.1 M KNO₃ added. (b) Linear sweep voltammograms of CuCo alloy in 1 M KOH containing varying concentrations of KNO₃. (c) FE and production rate of NO₃RR to ammonia over CuCo alloy in 1 M KOH with 0.1 M KNO₃ at a series of potentials. (d) FE of NO₃RR to NH₃ over Cu single metal, Co single metal, and CuCo alloy in 1 M KOH with 0.1 M KNO₃ at a series of potentials. (e) FE of NO₃RR to NO₂⁻ over Cu single metal, Co single metal, and CuCo alloy in 1 M KOH with 0.1 M KNO₃ at a series of potentials. (f)

Production Rate of NH$_3$ over Cu single metal, Co single metal, and CuCo alloy in 1 M KOH with 0.1 M KNO$_3$ at a series of potentials. (g) Double-layer capacitance per geometric area (C$_{dl}$) of Cu single metal, Co single metal, and CuCo alloy. (h) The It curves in the cycling tests of NO$_3$RR over CuNi DSACs at –0.425 V vs. RHE in 1 M KOH with 0.1 M KNO$_3$. (i) FE of NH$_3$ in the cycling tests of NO$_3$RR over CuNi DSACs at –0.425 V vs. RHE in 1 M KOH with 0.1 M KNO$_3$.

### 3.4 Insights into the reaction mechanism of CuCo Alloy towards NO$_3$RR

To elucidate the chemical states of the surface elements in the CuCo alloy, X-ray photoelectron spectroscopy (XPS) measurements were conducted using single-metal Cu and Co as references (**Figure S12**). In the CuCo alloy, the Co 2p spectrum presents principal doublets at binding energies of 778, 793, 779.4, 794.4, 782.2, and 797.2 eV[58,59], attributed to the respective oxidation states Co$^0$ 2$p_{3/2}$, Co$^0$ 2$p_{1/2}$, Co$^{3+}$ 2$p_{3/2}$, Co$^{3+}$ 2$p_{1/2}$, Co$^{2+}$ 2$p_{3/2}$, and Co$^{2+}$ 2$p_{1/2}$. When compared to the Co single metal, whose Co 2$p$ spectrum shows main doublets at slightly altered binding energies of 778, 793, 779.6, 794.6, 781.1, and 796.3 eV, correlating to the same oxidation states of Co as in the CuCo alloy, we observe shifts in binding energy, especially for Co$^{2+}$ and Co$^{3+}$. Upon comparative analysis of the Co 2$p$ XPS spectra between Co single metal and the CuCo alloy, it was observed that the binding energy for divalent Co$^{2+}$ in the alloy increased by 0.9 eV, while that for trivalent Co$^{3+}$ decreased by 0.2 eV. The elevation in binding energy indicates a reduction in electron density at the Co$^{2+}$ sites, whereas the decrease suggests an augmentation at the Co$^{3+}$ sites. These shifts reveal significant intermetallic interactions between Cu and Co during the alloying process, characterized by electron transfer leading to a redistribution of electron density. The formation of Co−Cu bonds in the alloy matrix introduces a modification in the electron density around the Co atoms, evidencing the strong electronic coupling between Cu and Co, a hallmark of the alloy's metallic bonding character.

Furthermore, the Cu 2$p$ spectrum of the CuCo alloy displays main doublets at binding energies of 931.9, 951.6, 933.2, and 953 eV, attributed to $Cu^0$ $2p_{3/2}$, $Cu^0$ $2p_{1/2}$, $Cu^{2+}$ $2p_{3/2}$, and $Cu^{2+}$ $2p_{1/2}$, respectively[60]. This contrasts with the Cu single metal, which exhibits principal doublets at slightly different binding energies of 932.5 and 953.2 eV, corresponding to $Cu^0$ $2p_{3/2}$ and $Cu^0$ $2p_{1/2}$. Notably, a reduction of 0.6 eV in the binding energy for $Cu^0$, alongside the detection of oxidized Cu states within the alloy, further corroborates the intricate electronic coupling between Cu and Co. This pronounced electron interplay during the Cu and Co alloying process not only facilitates the diversification of chemical states within the alloy but also significantly influences the alloy's electronic structure.

To elucidate the active phases involved in NO$_3$RR, we employed in situ Raman spectroscopy to monitor the evolution of phases across three catalysts subjected to various applied potentials in a solution of 1 M KOH and 0.1 M $NO_3^-$. The presence of $NO_3^-$ ions was confirmed by their distinctive peak at approximately 1050 cm$^{-1}$[61]. Additionally, K$_2$SO$_4$ was introduced to improve ionic conductivity and served as an external standard in Raman spectroscopy, providing $SO_4^{2-}$ ions that are marked by a typical signal at around 982 cm$^{-1}$[61].

**Figure 4**a shows the Raman spectra of Cu single metal at reducing potentials related to NO$_3$RR. In the NO$_3$RR process, the surface Cu species of the catalyst exhibit behaviors closely related to changes in reduction potentials. Notably, the characteristic peak of CuO at 425 cm$^{-1}$ remains relatively stable throughout the range of potentials examined. This persistent presence suggests that CuO may act as an important catalytic phase during NO$_3$RR, offering favorable sites for the adsorption and activation of nitrate, and participating in the crucial conversion step from $NO_3^-$ to $NO_2^-$. As the reduction potential increases, the characteristic peak of $Cu_2O$ emerges at 330 and 452 cm$^{-1}$[62], indicating the electrochemical reduction of Cu(II) to Cu(I). This transformation reflects the valence state changes of Cu species on the catalyst surface during the electrochemical reaction. Furthermore, as the nitrate reduction reaction proceeds, the consumption of protons in the reaction medium leads to an enrichment of $OH^-$. This shift promotes the formation of $Cu(OH)_2$ species, as evidenced by the characteristic peaks at 554 and 585 cm$^{-1}$ in the Raman spectrum[63]. Based on electrochemical data analysis (**Figure 3**e), the FE for $NO_2^-$ reduction decreases at higher voltages. This suggests that $Cu(OH)_2$ species may not only promote the further reduction of $NO_2^-$ to $NH_3$ but also indicate that, at more negative potentials, $Cu(OH)_2$ could act as an active intermediate in the catalytic reaction.

**Figure 4**b shows the Raman spectra of Co single metal at reducing potentials related to $NO_3RR$. In the exploration of $NO_3RR$ under strongly alkaline conditions, the surface of Co single metal catalysts at open circuit potential (OCP) displayed the phases of $Co(OH)_2$ and $Co_3O_4$. These phases are likely critical active sites for the adsorption and initial activation of nitrate. Raman spectroscopy revealed characteristic peaks at 425 and 545 cm$^{-1}$, which are attributed to $Co(OH)_2$[64], respectively, reflecting typical vibrations of the Co−O bonds within the layered structure of Co hydroxide and deformation vibrations of $OH^-$ ions. Additionally, the peak at 760 cm$^{-1}$ distinctly indicated the presence of the $Co_3O_4$ species, characteristic of Co−O bond vibrations within the Co spinel structure. As the potential became more negative, the original $Co(OH)_2$ signals gradually weakened, while new signals for the CoOOH species emerged at 590, 626 and 666 cm$^{-1}$.[65] This phenomenon contradicts the typical electrocatalytic expectation that catalysts shift from higher to lower oxidation states with increasing reduction potentials. This unusual behavior suggests that CoOOH may serve as an active site facilitating electron transfer in intermediary steps of $NO_3^-$ to NH$_3$ conversion during NO$_3$RR. With further increase in potential, the signals of Co$_3$O$_4$ and CoOOH diminished, implying the reduction of these higher oxidation states to metallic Co. Furthermore, according to **Figure 3**d and e, the FE of NH$_3$ and NO$_2^-$ decreased, indicating that HER became dominant at high reduction potentials. This likely reduced the number of active sites and inhibited the formation or persistence of Co$_3$O$_4$ and CoOOH on the catalyst surface, thus weakening or eliminating their signals.

**Figure 4**c shows the Raman spectra of CuCo alloy at reducing potentials related to NO$_3$RR. At OCP, the Raman spectrum predominantly shows a characteristic peak of CuO at 425 cm$^{-1}$, which overlaps with the peak of Co(OH)$_2$. This overlap is likely due to the similar vibrational characteristics of their metal−O bonds. Additionally, a signal for Co$_3$O$_4$ at 755 cm$^{-1}$ is observed, consistent with the behavior of Co single metal catalysts.

As the potential is adjusted to 0 V, a characteristic peak for Cu$_2$O appears at 330 cm$^{-1}$, indicating the reduction of CuO to Cu$_2$O. Concurrently, characteristic peaks for Cu(OH)$_2$ at 532, 566 cm$^{-1}$ emerge, suggesting further chemical transformation of the Cu species. At −0.1 V, the Raman signal of CoOOH was observed. By −0.2 V, the emergence of a new Cu$_2$O vibrational mode further confirms the ongoing transformation of the Cu species.

Compared to single-metal catalysts of Cu and Co, the majority of active phases in the CuCo alloy appear at lower reduction potentials and remain stable across a broader range of potentials. During the electrocatalytic NO$_3$RR, CuCo alloy, through the optimization of electron and atomic distribution, significantly improves the conversion efficiency of NO$_3^-$ to NH$_3$ while substantially reducing the generation of the byproduct NO$_2^-$. This synergistic effect reveals the complementary roles of Cu and Co sites in electron regulation and active site provision.

These properties further confirm that the active phases play a critical role in $NO_3RR$. Moreover, Raman spectroscopy of the CuCo alloy surface reveals that the $NO_3^-$ Raman peaks are weaker compared to those on single-metal catalysts, further suggesting the high efficiency of the CuCo alloy in $NO_3RR$.

Upon alloying, the electronic interactions between Cu and Co can lead to the formation of new electronic states, thereby enhancing the chemical and electrochemical stability of the alloy. These changes in electronic structure enable the alloy to maintain the stability of its active phases across a broader range of potentials. This new electronic configuration, resulting from the electron donation and acceptance mechanism between Cu and Co, differs from the electronic structure of the individual metals, thereby synergistically enhancing the catalytic activity and selectivity. Furthermore, the combination of Cu and Co also optimizes the adsorption characteristics on the catalyst surface. Cu enhances the adsorption of reaction intermediates, while Co assists in optimizing their chemical transformation processes. This synergistic interaction significantly reduces the activation energy of the reaction, accelerating the rate of chemical reactions (**Figure 4**d).

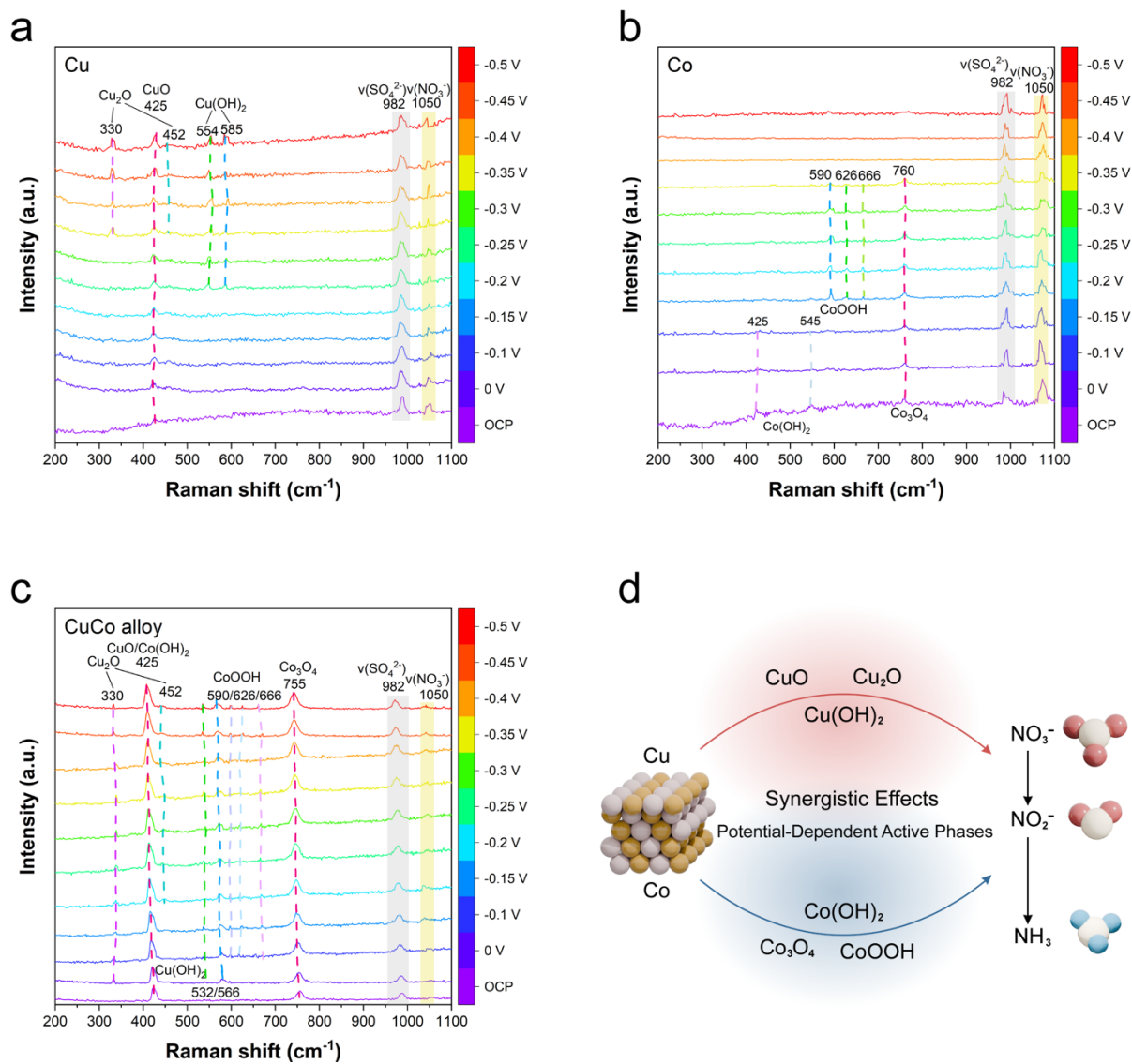

**Figure 4.** In situ Raman spectra of the catalysts and schematic of Synergistic Effects of CuCo alloy catalysts. In situ Raman spectra of Cu single metal (a), Co single metal (b) and CuCo alloy (c) at different applied potentials in electrolytes containing 0.1 M KNO$_3$, 0.1 M K$_2$SO$_4$ and 1 M KOH. (d) Reaction Mechanism of CuCo Alloy Synergistic Catalysis for NO$_3$RR. In the middle of (d) The light pink region corresponds to the active phase of Cu, while the light blue region

represents the active phase of Co, synergistically facilitate the reduction of $NO_3^-$ to $NO_2^-$, and subsequently $NO_2^-$ to $NH_3$.

### 3.5 The First Principal Study of CuCo Alloy

To further substantiate the origin of the catalytic activity within our CuCo alloy, we embarked on an in-depth investigation of the surface reactions during the nitrate reduction reaction ($NO_3RR$). Our research journey is encapsulated in **Figure 5**, where a suite of diagrams conveys the intricate dance of molecules and electrons across the catalytic interface.

In **Figure 5**a, we unveil a detailed scheme of the $NO_3RR$ as it unfolds on the CuCo alloy. This diagram, a distillation of countless hours of meticulous study, lays bare the stepwise transformation of $NO_3^-$ into $NH_3$. This visual representation is not merely illustrative but foundational to our understanding of the nuanced mechanisms at work. Therefore, in delving into the NO3RR on CuCo alloy, our analysis revealed distinctive behaviors of Cu and Co that underscored their individual limitations (**Figure 5**b). Cu, with its d-orbitals fully occupied, displayed a notable lack of affinity for $NO_3^-$, rendering it ineffective in capturing nitrate ions. On the other hand, Co, despite its ability to adsorb $NO_3^-$, formed strong bonds with $NH_3$, complicating the product's release and leading to potential catalyst poisoning by inhibiting further catalytic activity.

The creation of the CuCo alloy marked a turning point in our research, as evidenced in Figure 5(b). This alloy ingeniously combined the strengths of Cu and Co while overcoming their respective weaknesses. It facilitated efficient $NO_3^-$ adsorption—addressing Cu's shortfall—and allowed for the easy desorption of $NH_3$, avoiding the issue associated with Co. This balance resulted in a lowered energy barrier for the reaction, demonstrating the CuCo alloy's enhanced catalytic activity.

Through this innovative alloying approach, we achieved a synergistic effect that significantly improved the $NO_3RR$ process, making the CuCo alloy a superior catalyst.

To clarify the reaction process, we chart the comparative energy barriers and NH3 desorption energies of our compounds (**Figure 5**c). The energies we encountered told us a compelling story: when Co and Cu are united in a CuCo alloy, the resulting energy barrier is markedly reduced, facilitating a more accessible pathway for the reaction to proceed. The desorption energies we observed for $NH_3$ also spoke volumes about the synergistic effect of the CuCo alloy, which encourages the release of $NH_3$ without significant resistance.

This energy change can be illustrated by the surface charge density distribution of the CuCo alloy (**Figure 5**d). The color-coded map reveals the alloy's unique electronic landscape, where Cu and Co atoms alternate, creating a polarized surface. This distinctive feature is crucial, as it bestows upon the material the ability to act as both an electron donor and acceptor, a duality that is rare and highly beneficial for catalytic processes.

This unique atomic arrangement also causes the shifts in the d-band center of surface Cu and Co atoms upon the formation of the alloy (**Figure 5**e). Our findings are revealing: forming the CuCo alloy not only modifies the surface electronic structure but enhances it, elevating the catalytic prowess of the material. This increase in the d-band center is a clear indicator of a catalyst ready to engage more effectively with reactant molecules.

Conclusively, the CuCo alloy is not simply a merger of two metals, but a harmoniously integrated system that surpasses the capabilities of its individual components. This integration enables the alloy to address the adsorption-desorption conundrum, positioning it as a superior candidate for catalyzing the $NO_3RR$. Our findings not only advance the fundamental understanding of bimetallic catalysis but also pave the way for developing more efficient and environmentally friendly catalysts for nitrate reduction.

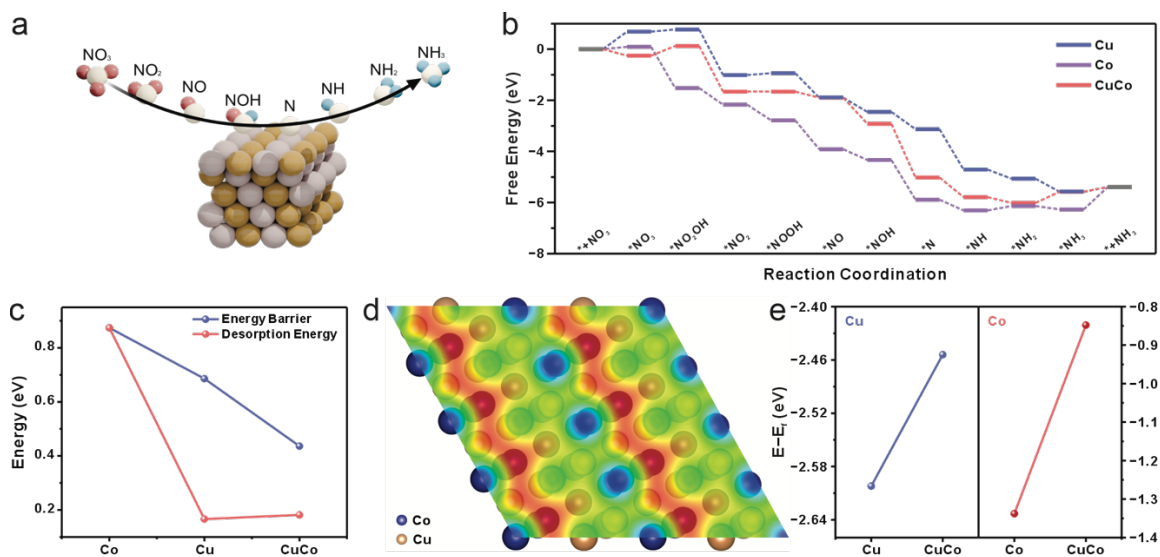

**Figure 5.** (a) Schematic representation of the NO$_3$RR pathway facilitated by a CuCo alloy catalyst. (b) Free energy profiles delineate the transformation sequence of intermediates during the NO$_3$RR on a CuCo alloy. (c) Correlated diagram contrasting the energy barriers with NH$_3$ desorption energies for an array of compounds. (d) Surface charge density distribution across the CuCo alloy, with an isosurface value of 0.0001 e/bohr$^3$ for visualization. (e) The d-band center alignments of Cu and Co atoms present on the surfaces of various metals, indicating electronic structural variations.

## 4 Conclusions

In summary, compared to traditional material synthesis exploration methods, this study effectively guided the selection of key experimental parameters using machine learning techniques and pioneered an entropy-engineered synthesis approach. This method successfully produced stable nanometric CuCo solid solution alloys. Notably, these alloys demonstrated excellent industrial-scale electrocatalytic performance in NO$_3$RR. The synergistic interaction mechanisms between the active phases of copper and cobalt were investigated using in situ Raman spectroscopy, with DFT calculations providing further validation of our findings.

ASSOCIATED CONTENT

**Supporting Information**. FE-SEM imaging, XRD patterns, Raman spectroscopy, and XPS analysis of CuCo alloys loaded on carbon nanofibers, highlighting the alloy's crystal structure, particle size distribution, and comparison with single metals Cu and Co. Additionally, electrocatalytic performance tests, particularly in $NO_3RR$, were conducted, and stability was assessed through FE-SEM and FE-TEM imaging. The results demonstrate the alloy's excellent stability and catalytic performance after repeated use.

AUTHOR INFORMATION

Corresponding Author

*Corresponding author. E-mail: lanhh@mit.edu, hcxia9209@zzu.edu.cn, du@jiangnan.edu.cn

ACKNOWLEDGMENT

REFERENCES

(1) (1) Liu, S.; Shen, Y.; Zhang, Y.; Cui, B.; Xi, S.; Zhang, J.; Xu, L.; Zhu, S.; Chen, Y.; Deng, Y.; Hu, W. Extreme Environmental Thermal Shock Induced Dislocation-Rich Pt Nanoparticles Boosting Hydrogen Evolution Reaction. *Adv. Mater.* **2022**, *34* (2), 2106973.

(2) Jia, L.; Sun, M.; Xu, J.; Zhao, X.; Zhou, R.; Pan, B.; Wang, L.; Han, N.; Huang, B.; Li, Y. Phase-Dependent Electrocatalytic $CO_2$ Reduction on $Pd_3Bi$ Nanocrystals. *Angew. Chem. Int. Ed.* **2021**, *133* (40), 21909–21913.

(3) Hu, Y.; khan, M. K.; Gong, J.; Zeb, H.; Lan, H.; Asif, M.; Xia, H.; Du, M. Mechanistic Insights into C–C Coupling in Electrocatalytic $CO_2$ Reduction Reaction. *Chem. Commun.* **2024**. https://doi.org/10.1039/D4CC03964E.

(4) Cui, B.; Zhu, H.; Wang, M.; Zeng, J.; Zhang, J.; Tian, Z.; Jiang, C.; Sun, Z.; Yang, H.; Liu, Y.; Ding, J.; Luo, Z.; Chen, Y.; Chen, W.; Hu, W. Intermediate State of Dense Ru Assembly Captured by High-Temperature Shock for Durable Ampere-Level Hydrogen Production. *ACS Mater. Lett.* **2024**, *6* (4), 1532–1541.

(5) Xin, C.; Wang, N.; Chen, Y.; He, B.; Zhao, Q.; Chen, L.; Tang, Y.; Luo, B.; Zhao, Y.; Yang, X. Biological Corrosion Behaviour and Antibacterial Properties of Ti-Cu Alloy with Different $Ti_2Cu$ Morphologies for Dental Applications. *Mater. Des.* **2022**, *215*, 110540.

(6) Shi, Z.; Li, Y.; Wu, X.; Zhang, K.; Gu, J.; Sun, W.; Li, C. M.; Guo, C. X. Graphdiyne Chelated AuNPs for Ultrasensitive Electrochemical Detection of Tyrosine. *Chem. Commun.* **2023**, *59* (91), 13647–13650.

(7) Cao, G.; Liang, J.; Guo, Z.; Yang, K.; Wang, G.; Wang, H.; Wan, X.; Li, Z.; Bai, Y.; Zhang, Y.; Liu, J.; Feng, Y.; Zheng, Z.; Lu, C.; He, G.; Xiong, Z.; Liu, Z.; Chen, S.; Guo, Y.; Zeng, M.; Lin, J.; Fu, L. Liquid Metal for High-Entropy Alloy Nanoparticles Synthesis. *Nature* **2023**, *619* (7968), 73–77.

(8) Lin, F.; Li, M.; Zeng, L.; Luo, M.; Guo, S. Intermetallic Nanocrystals for Fuel-Cells-Based Electrocatalysis. *Chem. Rev.* **2023**, *123* (22), 12507–12593.

(9) Hu, Y.; Lan, H.; He, J.; Fang, W.; Zhang, W.-D.; Lu, S.; Duan, F.; Du, M. Entropy-Engineered Middle-In Synthesis of Dual Single-Atom Compounds for Nitrate Reduction Reaction. *ACS Nano* **2024**. https://doi.org/10.1021/acsnano.4c05568.

(10) Yao, Y.; Huang, Z.; Xie, P.; Lacey, S. D.; Jacob, R. J.; Xie, H.; Chen, F.; Nie, A.; Pu, T.; Rehwoldt, M.; Yu, D.; Zachariah, M. R.; Wang, C.; Shahbazian-Yassar, R.; Li, J.; Hu, L. Carbothermal Shock Synthesis of High-Entropy-Alloy Nanoparticles. *Science* **2018**, *359* (6383), 1489–1494.

(11) Calvo, F. The Interplay between Size, Shape, and Surface Segregation in High-Entropy Nanoalloys. *Phys. Chem. Chem. Phys.* **2023**, *25* (27), 18439–18453.

(12) Sun, J.; Leff, A.; Li, Y.; Woehl, T. J. Visualizing Formation of High Entropy Alloy Nanoparticles by Aggregation of Amorphous Metal Cluster Intermediates *Nanoscale* **2023**, 15, 10447-10457.

(13) Lan, H.; Wang, L.; He, R.; Huang, S.; Yu, J.; Guo, J.; Luo, J.; Li, Y.; Zhang, J.; Lin, J.; Zhang, S.; Zeng, M.; Fu, L. 2D Quasi-Layered Material with Domino Structure. *Nat. Commun.* **2023**, *14* (1), 7225.

(14) Qi, W.; Wang, M.; Su, Y. Size Effect on the Lattice Parameters of Nanoparticles. *J. Mater. Sci. Lett.* **2002**, *21*, 877–878.

(15) Ludwig, A. Maximize Mixing in Highly Polyelemental Solid Solution Alloy Nanoparticles. *Matter* **2021**, *4* (7), 2100–2101.

(16) Liu, Y.; Yang, Z.; Yu, Z.; Liu, Z.; Liu, D.; Lin, H.; Li, M.; Ma, S.; Avdeev, M.; Shi, S. Generative Artificial Intelligence and Its Applications in Materials Science: Current Situation and Future Perspectives. *J. Mater. Sci. Lett.* **2023**, *9* (4), 798–816.

(17) Mobarak, M. H.; Mimona, M. A.; Islam, Md. A.; Hossain, N.; Zohura, F. T.; Imtiaz, I.; Rimon, M. I. H. Scope of Machine Learning in Materials Research—A Review. *Appl. Surf. Sci. Adv.* **2023**, *18*, 100523.

(18) Valizadeh, A.; Sahara, R.; Souissi, M. Alloys Innovation through Machine Learning: A Statistical Literature Review. *Sci. Technol. Adv. Mater. Methods* **2024**, *4* (1), 2326305.

(19) Zhichao, L.; Dong, M.; Xiongjun, L.; Lu, Z. High-Throughput and Data-Driven Machine Learning Techniques for Discovering High-Entropy Alloys. *Commun. Mater.* **2024**, *5* (1), 1–19.

(20) Hong, Y.; Hou, B.; Jiang, H.; Zhang, J. Machine Learning and Artificial Neural Network Accelerated Computational Discoveries in Materials Science. *WIREs Comput. Mol. Sci.* **2020**, *10* (3), e1450.

(21) He, W.; Zhang, J.; Dieckhöfer, S.; Varhade, S.; Brix, A. C.; Lielpetere, A.; Seisel, S.; Junqueira, J. R. C.; Schuhmann, W. Splicing the Active Phases of Copper/Cobalt-Based Catalysts Achieves High-Rate Tandem Electroreduction of Nitrate to Ammonia. *Nat. Commun.* **2022**, *13* (1), 1129.

(22) Wang, Y.; Zhou, W.; Jia, R.; Yu, Y.; Zhang, B. Unveiling the Activity Origin of a Copper-based Electrocatalyst for Selective Nitrate Reduction to Ammonia. *Angew. Chem. Int. Ed.* **2020**, *59* (13), 5350–5354.

(23) Hodgetts, R. Y.; Kiryutin, A. S.; Nichols, P.; Du, H.-L.; Bakker, J. M.; Macfarlane, D. R.; Simonov, A. N. Refining Universal Procedures for Ammonium Quantification via Rapid $^1$H NMR Analysis for Dinitrogen Reduction Studies. *ACS Energy Lett.* **2020**, *5* (3), 736–741.

(24) Kresse, G.; Furthmüller, J. Efficient Iterative Schemes for Ab Initio Total-Energy Calculations Using a Plane-Wave Basis Set. *Phys. Rev. B* **1996**, *54* (16), 11169–11186.

(25) Blöchl, P. E. Projector Augmented-Wave Method. *Phys. Rev. B* **1994**, *50* (24), 17953–17979.

(26) Wei, Y.; Xia, H.; Lan, H.; Xue, D.; Zhao, B.; Yu, Y.; Hu, Y.; Zhang, J. Boosting the Catalytic Activity of Nitrogen Sites by Spin Polarization Engineering for Oxygen Reduction and Wide-Temperature Ranged Quasi-Solid Zn–Air Batteries. *Adv. Energy Mater.* **2024**, *14* (4), 1–11.

(27) Li, L.; Dang, W.; Zhu, X.; Lan, H.; Ding, Y.; Li, Z.; Wang, L.; Yang, Y.; Fu, L.; Miao, F.; Zeng, M. Ultrathin Van Der Waals Lanthanum Oxychloride Dielectric for 2D Field-Effect Transistors. *Adv. Mater.* **2023**, *2309296*, 1–8.

(28) Lan, H.; Li, Y.; Liu, J.; Hu, W.; Zhu, X.; Ma, Y.; Niu, L.; Zhang, Z.; Jia, S.; Li, L.; Chen, Y.; Wang, J.; Zeng, M.; Fu, L. Self-Limiting Synthesis of Ultrathin Ge(110) Single Crystal via Liquid Metal. *Small* **2022**, *18* (9), 1–7.

(29) Lan, H.; Wang, L.; Li, Y.; Deng, S.; Yue, Y.; Zhang, T.; Zhang, S.; Zeng, M.; Fu, L. Self-Modulation-Guided Growth of 2D Tellurides with Ultralow Thermal Conductivity. *Small* **2022**, *18* (41), 1–8.

(30) Shen, Y.; Zou, J.; Lan, H.; Ding, Y.; Liang, Z.; Yang, Z.; Zeng, Z.; Long, J.; Zhao, Y.; Fu, L.; Zeng, M. Unlocking Prussian Blue Analogues Inert-Site to Achieve High-Capacity Ammonium Storage. *Adv. Funct. Mater.* **2024**, *2400598*, 1–10.

(31) Wu, Y.; Zhang, Y. Design and High-Throughput Screening of High Entropy Alloys. In *Advances in High-Entropy Alloys - Materials Research, Exotic Properties and Applications*; Kitagawa, J., Ed.; IntechOpen: Rijeka **2021**, p Ch. 6.

(32) Wang, B.; Lei, S.; Li, C.; Gao, A. Effect of Ambient Atmosphere on the Formation and Evolution of Conjugated Structure of Pre-Oxidized Polyacrylonitrile. *Mater. Res. Express* **2022**, *9* (5), 55601.

(33) Samimi-Sohrforozani, E.; Azimi, S.; Abolhasani, A.; Malekian, S.; Arbab, S.; Zendehdel, M.; Abolhasani, M. M.; Yaghoobi Nia, N. Development of Porous Polyacrylonitrile Composite Fibers: New Precursor Fibers with High Thermal Stability. *Electron. Mater.* **2021**, pp 454–465.

(34) Bolar, S.; Ito, Y.; Fujita, T. Future Prospects of High-Entropy Alloys as Next-Generation Industrial Electrode Materials. *Chem. Sci.* **2024**, 8664–8722.

(35) Bokas, G. B.; Chen, W.; Hilhorst, A.; Jacques, P. J.; Gorsse, S.; Hautier, G. Unveiling the Thermodynamic Driving Forces for High Entropy Alloys Formation through Big Data Ab Initio Analysis. *Scr. Mater.* **2021**, *202*, 114000.

(36) Evans, D.; Chen, J.; Bokas, G.; Chen, W.; Hautier, G.; Sun, W. Visualizing Temperature-Dependent Phase Stability in High Entropy Alloys. *NPJ Comput. Mater.* **2021**, *7* (1), 1–9.

(37) Melnick, O. B.; Soolshenko, V. K.; Levchuk, K. H. Thermodynamic Prediction of Phase Composition of Transition Metals High-Entropy Alloys. *Metallofiz. Noveishie Tekhnol.* **2020**, *42* (10), 1387–1400.

(38) Carlucci, G.; Motta, C.; Casati, R. High-Throughput Design of Refractory High-Entropy Alloys: Critical Assessment of Empirical Criteria and Proposal of Novel Guidelines for Prediction of Solid Solution Stability. *Adv. Eng. Mater.* **2024**, *26* (3), 2301425.

(39) V, B.; M, A. X. Development of High Entropy Alloys (HEAs): Current Trends. *Heliyon* **2024**, *10* (7), e26464.

(40) Sabooni, S.; Karimzadeh, F.; Abbasi, M. H. Thermodynamic Aspects of Nanostructured Ti 5Si 3 Formation during Mechanical Alloying and Its Characterization. *Bull. Mater. Sci.* **2012**, *35* (3), 439–447.

(41) Moazzen, P.; Toroghinejad, M. R.; Karimzadeh, F.; Vleugels, J.; Ravash, H.; Cavaliere, P. Influence of Zirconium Addition on the Microstructure, Thermodynamic Stability, Thermal Stability and Mechanical Properties of Mechanical Alloyed Spark Plasma Sintered (MA-SPS) FeCoCrNi High Entropy Alloy. *Powder Metall.* **2018**, *61* (5), 405–416.

(42) Zhao, X.; Kong, X.; Li, G.; Zhao, Y.; Jia, Z.; He, F.; Yang, P.; Ge, K.; Zhang, M.; Liu, Z. Ru-Based Catalysts for Hydrogenation of N-Ethylcarbazole: Progress and Prospects. *Fuel* **2024**, *360*, 130605.

(43) Ma, T.; Wang, W.; Wang, R. Thermal Degradation and Carbonization Mechanism of Fe-Based Metal-Organic Frameworks onto Flame-Retardant Polyethylene Terephthalate. *Polymers*. **2023**, 15(1), 224.

(44) Yeo, L. Y.; Friend, J. R. Electrospinning Carbon Nanotube Polymer Composite Nanofibers. *J. Exp. Nanosci.* **2006**, *1* (2), 177–209.

(45) Zhang, Z.; Wang, Z.; He, S.; Wang, C.; Jin, M.; Yin, Y. Redox Reaction Induced Ostwald Ripening for Size- and Shape-Focusing of Palladium Nanocrystals. *Chem. Sci.* **2015**, *6* (9), 5197–5203.

(46) Yao, Y.; Huang, Z.; Hughes, L. A.; Gao, J.; Li, T.; Morris, D.; Zeltmann, S. E.; Savitzky, B. H.; Ophus, C.; Finfrock, Y. Z.; Dong, Q.; Jiao, M.; Mao, Y.; Chi, M.; Zhang, P.; Li, J.; Minor, A. M.; Shahbazian-Yassar, R.; Hu, L. Extreme Mixing in Nanoscale Transition Metal Alloys. *Matter* **2021**, *4* (7), 2340–2353.

(47) Abdelhafiz, A.; Wang, B.; Harutyunyan, A. R.; Li, J. Carbothermal Shock Synthesis of High Entropy Oxide Catalysts: Dynamic Structural and Chemical Reconstruction Boosting the Catalytic Activity and Stability toward Oxygen Evolution Reaction. *Adv. Energy Mater.* **2022**, *12* (35), 10–15.

(48) Buckingham, M. A.; Skelton, J. M.; Lewis, D. J. Synthetic Strategies toward High Entropy Materials: Atoms-to-Lattices for Maximum Disorder. *Cryst. Growth Des.* **2023**, *23* (10), 6998–7009.

(49) Nie, J.; Zhi, Y.; Fan, Y.; Zhao, Y.; Liu, X. One-Step Synthesis of High-Entropy Diborides with Hierarchy Structure and High Hardness via Aluminum-Melt Reaction Method. *Mater. Res. Lett.* **2024**, *12* (2), 88–96.

(50) Han, Y.-C.; Cao, P.-Y.; Tian, Z.-Q. Controllable Synthesis of Solid Catalysts by High-Temperature Pulse. *Acc. Mater. Res.* **2023**, *4* (8), 648–654.

(51) Kozak, R.; Sologubenko, A.; Steurer, W. Single-Phase High-Entropy Alloys - An Overview. *Z. Kristallogr.* **2015**, *230* (1), 55–68.

(52) Zhang, R. Z.; Reece, M. J. Review of High Entropy Ceramics: Design, Synthesis, Structure and Properties. *J. Mater. Chem.* **2019**, *7* (39), 22148–22162.

(53) Rogachev, A. S. Structure, Stability, and Properties of High-Entropy Alloys. *Phys. Met. Metallogr.* **2020**, *121*, 733+.

(54) Peter, N. J.; Duarte, M. J.; Liebscher, C. H.; Srivastava, V. C.; Uhlenwinkel, V.; Jägle, E. A.; Dehm, G. Early Stage Phase Separation of AlCoCr0.75Cu0.5FeNi High-Entropy Powder at the Nanoscale. *J. Alloys Compd.* **2020**, *820*, 153149.

(55) Lucia, U.; Grazzini, G. The Second Law Today: Using Maximum-Minimum Entropy Generation. *Entropy* **2015**, *17* (11), 7786–7797.

(56) Bulut, A.; Yurderi, M.; Ertas, İ. E.; Celebi, M.; Kaya, M.; Zahmakiran, M. Carbon Dispersed Copper-Cobalt Alloy Nanoparticles: A Cost-Effective Heterogeneous Catalyst with Exceptional Performance in the Hydrolytic Dehydrogenation of Ammonia-Borane. *Appl. Catal. B* **2016**, *180*, 121–129.

(57) Hu, H.; Miao, R.; Yang, F.; Duan, F.; Zhu, H.; Hu, Y.; Du, M.; Lu, S. Intrinsic Activity of Metalized Porphyrin-based Covalent Organic Frameworks for Electrocatalytic Nitrate Reduction. *Adv. Energy Mater.* **2024**, *14* (6), 1–10.

(58) Fu, Y.; Yu, H.-Y.; Jiang, C.; Zhang, T.-H.; Zhan, R.; Li, X.; Li, J.-F.; Tian, J.-H.; Yang, R. NiCo Alloy Nanoparticles Decorated on N-Doped Carbon Nanofibers as Highly Active and Durable Oxygen Electrocatalyst. *Adv. Funct. Mater.* 2018, *28* (9), 1705094.

(59) He, C.; Wang, S.; Jiang, X.; Hu, Q.; Yang, H.; He, C. Bimetallic Cobalt–Copper Nanoparticle-Decorated Hollow Carbon Nanofibers for Efficient CO2 Electroreduction. *Front. Chem.* **2022**, *10*, 1–8.

(60) Liu, W.; Wang, J.; Cai, N.; Wang, J.; Shen, L.; Shi, H.; Yang, D.; Feng, X.; Yu, F. Porous Carbon Nanofibers Loaded with Copper-Cobalt Bimetallic Particles for Heterogeneously Catalyzing Peroxymonosulfate to Degrade Organic Dyes. *J. Environ. Chem. Eng.* **2021**, *9* (5), 106003.

(61) He, W.; Zhang, J.; Dieckhöfer, S.; Varhade, S.; Brix, A. C.; Lielpetere, A.; Seisel, S.; Junqueira, J. R. C.; Schuhmann, W. Splicing the Active Phases of Copper/Cobalt-Based Catalysts Achieves High-Rate Tandem Electroreduction of Nitrate to Ammonia. *Nat. Commun.* **2022**, *13* (1), 1129.

(62) Zhao, Y.; Chang, X.; Malkani, A. S.; Yang, X.; Thompson, L.; Jiao, F.; Xu, B. Speciation of Cu Surfaces During the Electrochemical CO Reduction Reaction. *J. Am. Chem. Soc.* **2020**, *142* (21), 9735–9743.

(63) Huang, K.; Tang, K.; Wang, M.; Wang, Y.; Jiang, T.; Wu, M. Boosting Nitrate to Ammonia via the Optimization of Key Intermediate Processes by Low-Coordinated Cu–Cu Sites. *Adv. Funct. Mater.* **2024**, *34* (24), 2315324.

(64) Yang, J.; Liu, H.; Martens, W. N.; Frost, R. L. Synthesis and Characterization of Cobalt Hydroxide, Cobalt Oxyhydroxide, and Cobalt Oxide Nanodiscs. *J. Phys. Chem. C* **2010**, *114* (1), 111–119.

(65) Liu, Y.-C.; Koza, J. A.; Switzer, J. A. Conversion of Electrodeposited Co(OH)2 to CoOOH and Co3O4, and Comparison of Their Catalytic Activity for the Oxygen Evolution Reaction. *Electrochim. Acta* **2014**, *140*, 359–365.

Table of Contents graphic

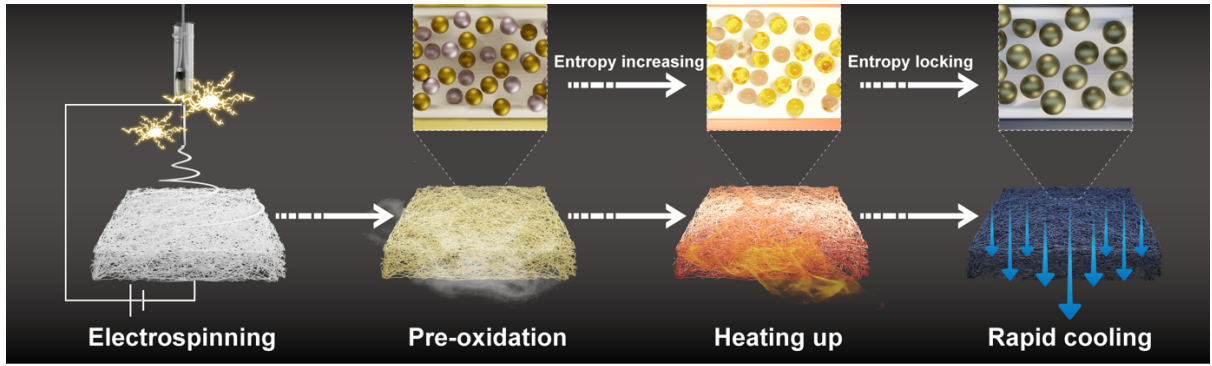